\documentclass[10pt]{article}
\usepackage[margin=1in]{geometry}
\usepackage{booktabs}
\usepackage{float}
\usepackage[hidelinks]{hyperref}

\title{Double KV Cache Optimization on Apple Silicon:\\
Block-FP4, Radial Quantization, and Fused Metal Kernels}
\author{%
Abduchaborov Ibrokhimsho\\
\small Independent Researcher, Tajikistan\\
\small \href{mailto:ibrokhimsho.a@gmail.com}{ibrokhimsho.a@gmail.com}\\
\small \href{https://orcid.org/0009-0009-2287-5266}{ORCID: 0009-0009-2287-5266}%
}
\date{\today}

\begin{document}
\maketitle

\begin{abstract}
Key--value (KV) cache capacity and memory traffic are primary bottlenecks in
long-context autoregressive inference \cite{pope2023efficiently,shazeer2019mqa}.
We evaluate five cache paths on an Apple Silicon M5 Max using MLX: a native
bf16 baseline, block-scaled E2M1 FP4, a Python reference implementing radial
quantization with a QJL-style residual, sequential FP4-to-Polar composition,
and a fused Metal Polar kernel. On Apple Silicon M5 Max, fused Metal PolarQuant
achieves 46.8 tok/s at 3,456 B/token---3.64$\times$ compression versus the bf16
baseline and 1.84$\times$ faster than the Python Polar reference (25.4 tok/s),
while slightly outperforming block-FP4 (41.5 tok/s at 3.62$\times$
compression). Sequential composition of FP4 and Polar (Double Optimization)
yields no additive benefit: 4,241 B/token at 15.1 tok/s, worse than FP4 alone
on both axes. The fused path establishes that radial coding can be competitive
when packing, rotation, and reconstruction are placed in one device kernel,
but this result does not extend to longer contexts. At 32K and above, all
compressed variants collapse to 1.5--14 tok/s because the current attention
path repeatedly reconstructs the full cache, retaining cumulative
$O(N^2)$ decode work. At 64K, the Metal variant runs out of memory when a
33.5 GB allocation exceeds MLX's 22.6 GB Metal buffer limit. Fused scaled
dot-product attention operating directly on compressed KV, together with
chunked pre-allocation, is required for long-context viability. These results
separate representation-level compression from kernel and allocator effects
and identify the conditions under which low-bit KV caching helps on unified
memory hardware.
\end{abstract}

\section{Introduction}
Autoregressive Transformer inference reuses the key and value tensors produced
for earlier tokens. This KV cache avoids recomputing the entire prefix at every
generation step, but its capacity grows linearly with sequence length and batch
size. The cache is read repeatedly during decoding, so its footprint is also a
bandwidth cost rather than merely a storage cost. Analytical and empirical
studies of Transformer serving identify this state as a central constraint on
long-context inference \cite{pope2023efficiently}. Multi-query attention
reduces the number of key/value heads \cite{shazeer2019mqa}, and grouped-query
attention interpolates between multi-head and multi-query designs
\cite{ainslie2023gqa}, yet large contexts still leave substantial persistent
state.

KV compression addresses the remaining state without changing model weights.
The design space includes scalar and group-wise quantization, sparse outlier
retention, token eviction, cross-layer sharing, geometric transforms, and
compressed-domain attention. These approaches make different tradeoffs. A
compact representation can save memory but lose throughput if every decode
step reconstructs the entire cache. Conversely, a specialized kernel can make
a less compact layout faster by avoiding framework dispatch, intermediate
tensors, and repeated format conversion. The representation and the consuming
attention kernel therefore have to be evaluated together.

Apple Silicon is a useful target for this comparison. CPU and GPU allocations
share physical memory, MLX schedules lazy array graphs onto Metal, and the
device exposes large unified-memory capacity without a discrete-host transfer.
Those properties remove one class of PCIe bottlenecks while exposing allocator
behavior and kernel granularity more directly. They also make it possible to
test whether a custom Metal implementation changes the practical ranking of KV
formats that look similar from byte accounting alone.

We study a 4-bit block representation, a radial representation motivated by
PolarQuant \cite{han2025polarquant}, a QJL-style residual
\cite{zandieh2024qjl}, their sequential composition, and a fused Metal radial
kernel. The baseline is the model's native bf16 cache. All reported numbers use
the same M5 Max system and fixed context configurations. Cache bytes are
normalized by the effective number of cached tokens; decode throughput is
measured after warmup.

The main findings are as follows.
\begin{itemize}
  \item At 8K, fused Metal Polar reaches 46.8 tok/s and 3,456 B/token. It is
  1.84$\times$ faster than Python Polar, 13\% faster than block-FP4, and nearly
  identical to block-FP4 in compression.
  \item The 8K bf16 baseline remains fastest at 171.4 tok/s. Compression saves
  capacity but does not automatically accelerate decode when the cache is
  reconstructed before standard attention.
  \item Sequential FP4-to-Polar composition is Pareto-dominated: 3,473 B/token
  and 41.5 tok/s for FP4 alone versus 4,241 B/token and 15.1 tok/s for the
  double path.
  \item At 32K, compressed decode falls to 1.5--14 tok/s while the baseline
  reaches 91.8 tok/s. Fusion reduces a constant factor at 8K but does not
  change the current algorithm's cumulative $O(N^2)$ work.
  \item At 64K, Python Polar completes at 0.47 tok/s and 3,072 B/token, whereas
  fused Metal Polar requests a 33.5 GB buffer and exceeds MLX's 22.6 GB
  per-buffer limit.
\end{itemize}

These results provide both a positive kernel result and two negative system
results. Fusion makes radial quantization competitive at moderate context, but
codec stacking does not yield multiplicative savings, and whole-cache
reconstruction prevents the compressed layouts from remaining competitive at
long context. The paper consequently treats bytes/token, decode throughput,
prefill throughput, and allocation failure as separate outcomes.

\section{Background and Related Work}
\subsection{KV cache growth and serving}
The Transformer computes attention from queries against earlier keys and
values \cite{vaswani2017attention}. In incremental decoding, each new token
adds cache entries and then attends over the retained prefix. If the context
length is $N$, the persistent cache capacity is $O(N)$, the attention work for
one late decode step is $O(N)$, and cumulative work across an $O(N)$ decode
horizon is $O(N^2)$. The distinction matters here: quantization reduces the
linear storage term but cannot remove the attention scan unless the attention
algorithm itself changes.

PagedAttention reduces fragmentation and enables flexible sharing across
requests \cite{kwon2023pagedattention}. FlashAttention reduces memory traffic
through IO-aware tiling \cite{dao2022flashattention}. These methods optimize
where KV lives or how exact attention consumes it. Our implementation instead
changes the representation of persistent KV and measures the cost of returning
it to an attention-compatible form.

Architectural methods reduce the amount of KV produced. Multi-query attention
shares one key/value head across query heads \cite{shazeer2019mqa}, while GQA
uses an intermediate number of KV heads \cite{ainslie2023gqa}. Such choices are
orthogonal to element precision: a GQA cache can still be represented in bf16,
FP4, or a radial code. This work holds the model architecture fixed so that
differences arise from the cache path rather than from retraining or changing
the number of KV heads.

\subsection{Low-bit quantization}
Weight quantization methods show that low precision can be practical when
scaling, outliers, and hardware execution are designed together. SmoothQuant
migrates activation difficulty into weights \cite{xiao2022smoothquant}; GPTQ
uses approximate second-order information for one-shot weight compression
\cite{frantar2022gptq}. KV tensors differ because they are generated online and
read after every token.

KIVI finds an asymmetry between keys and values and applies per-channel key and
per-token value quantization \cite{liu2024kivi}. KVQuant combines pre-RoPE
quantization, non-uniform codebooks, and sparse outlier handling
\cite{hooper2024kvquant}. GEAR combines quantization with low-rank and sparse
error corrections \cite{kang2024gear}. ZipCache and MiKV assign precision
according to token salience \cite{he2024zipcache,yang2024mikv}. These methods
illustrate why a nominal bit-width alone is insufficient: metadata, retained
outliers, and layout alignment all contribute to actual bytes/token.

\subsection{Radial coding and compressed-domain estimation}
QJL uses a Johnson--Lindenstrauss transform followed by sign quantization to
estimate inner products without conventional scale and zero-point metadata
\cite{zandieh2024qjl}. PolarQuant applies randomized preconditioning and a
polar transform so that angular coordinates follow a concentrated distribution
that is amenable to quantization \cite{han2025polarquant}. TurboQuant extends
the geometric view with online vector quantization and a near-optimal
distortion analysis \cite{zandieh2025turboquant}.

The present Python Polar path follows this family at the representation level
but reconstructs tensors before standard attention. It is therefore not a
compressed-domain estimator. The fused Metal path combines encode/decode
operations into a custom device kernel, but it also stops short of fused
scaled dot-product attention (SDPA). This boundary explains the 8K improvement
and the 32K collapse: kernel fusion removes launch and intermediate overhead,
yet the full reconstructed cache is still materialized and scanned.

\subsection{Selection, reuse, and cross-layer sharing}
Token-selection methods reduce the retained sequence rather than the bytes per
retained element. H$_2$O keeps recent tokens and heavy hitters
\cite{zhang2023h2o}; SnapKV selects clustered prompt positions
\cite{li2024snapkv}; PyramidKV varies the cache budget by layer
\cite{cai2024pyramidkv}; StreamingLLM retains attention sinks for bounded
streaming \cite{xiao2023streamingllm}. CacheGen compresses cache state for
transfer and reuse \cite{liu2023cachegen}. These approaches can be composed
with low-bit storage, but doing so changes retention semantics and requires
separate quality evaluation.

DeepSeek-V4.1-Flash demonstrates the scale of architectural composition. Its
reported 890 B/token global KV footprint combines FP4 storage with a Causal
Encoder--Decoder (CED) design and cross-layer reuse in Compressed Sparse
Attention 2 (CSA2) \cite{deepseek2026v41flash}. The full result is not an
FP4-only baseline. Our comparison isolates the quantization axis and uses the
DeepSeek result to identify, rather than obscure, the remaining architectural
gap.

\section{Cache Implementations}
\subsection{Native bf16 baseline}
The baseline uses the cache representation selected by MLX-LM for the loaded
model and stores key/value activations in bf16. It is not an unoptimized Python
reference: it benefits from MLX's standard execution path and therefore
provides the system throughput that a compressed implementation must approach.
The measured cache footprint is 12,562 B/token at 8K, 12,359 B/token at 32K,
and 12,323 B/token at 64K. Small variation across contexts comes from fixed
layout components and the effective cache offset.

\subsection{Block-scaled E2M1 FP4}
The block-FP4 codec groups 16 channels, computes one E4M3-like scale for the
group, maps normalized values to signed E2M1 symbols, and packs two 4-bit
symbols into each byte. The persistent representation includes both packed
nibbles and scale metadata. Attention consumes reconstructed values in the
model's activation type.

At 8K, the implementation occupies 3,473 B/token, corresponding to
3.62$\times$ compression from the 12,562 B/token baseline. It decodes at
41.5 tok/s versus 171.4 tok/s for bf16. At 32K, it occupies 3,460 B/token and
reaches 14.0 tok/s. These values show that the codec's storage ratio is stable
across context while its runtime cost grows with the amount of reconstructed
state.

The implementation is a structural simulation of a block-scaled FP4 cache. It
does not claim to reproduce a proprietary kernel or a complete DeepSeek
architecture. Its importance in this study is that the FP4-only footprint at
8K, 3,473 B/token, closely matches the approximately 3,460 B/token FP4-only
ablation associated with the DeepSeek comparison, providing a sanity check on
the layout accounting.

\subsection{Python Polar reference}
The Python path applies a seeded Walsh--Hadamard preconditioner and groups
adjacent rotated coordinates into polar pairs. Radius and angle are quantized,
and a QJL-style residual sign sketch can be retained. This representation
tests the geometric codec while leaving array orchestration to MLX and Python.
It therefore exposes dispatch, graph construction, temporary buffers, and
reconstruction overhead.

The canonical paper measurements use 4,241 B/token at 8K and 4,228 B/token at
32K. Decode throughput is 25.4 tok/s and 1.83 tok/s, respectively. The 64K path
uses the later compact 3,072 B/token layout and reaches 0.47 tok/s. Because
the 64K run uses the compact layout, its bytes/token are not directly identical
to the earlier 8K/32K Python layout; the table reports the measured value rather
than projecting one layout across contexts.

\subsection{Sequential FP4-to-Polar composition}
The double path first reconstructs the block-FP4 representation and then
applies radial coding. Only the final polar representation is retained. A
naive expectation is that the compression factors multiply. That expectation
is incorrect because the two codecs operate on incompatible distributions and
the intermediate FP4 representation is not independently stored.

At 8K, the double path occupies 4,241 B/token---the same as the Python Polar
layout and worse than FP4's 3,473 B/token. Decode reaches 15.1 tok/s, compared
with 41.5 tok/s for FP4. At 32K it occupies 4,228 B/token and reaches
3.5 tok/s. The path adds transform work while retaining the larger final
format, producing no additive capacity benefit.

\subsection{Fused Metal Polar kernel}
The fused path moves radial packing and reconstruction into a Metal kernel so
that operations that previously created separate MLX arrays execute in one
device program. Its layout occupies exactly 3,456 B/token at both 8K and 32K.
Relative to the 8K baseline, 3,456 B/token is 3.64$\times$ compression and is
slightly more compact than the 3,473 B/token FP4 path.

Fusion improves 8K decode from 25.4 tok/s for the Python Polar reference to
46.8 tok/s. This is the central positive systems result: the radial
representation is not intrinsically slower than block-FP4 at moderate context.
The kernel boundary, however, ends before SDPA. The implementation reconstructs
the cache for attention and updates storage through an iterative concatenation
path. Those choices dominate at 32K and trigger the 64K allocation failure.

\section{Experimental Methodology}
\subsection{Hardware and software}
Experiments run on a MacBook Pro with an Apple M5 Max, a 32-core integrated
GPU, and 36 GB of unified memory. The operating system is macOS 27.0.1. The
benchmark environment uses Python 3.14.5 and MLX-LM 0.32.2. The target model is
\texttt{mlx-community/LFM2.5-8B-A1B-OptiQ-4bit}. Weight quantization remains
fixed across all cache conditions; only KV handling changes.

Each context is constructed from deterministic token IDs. The requested
lengths are 8,192, 32,768, and 65,536 tokens; model framing produces effective
lengths close to these requests. Decode uses a fixed 64-token measured window
after warmup for the fused-kernel study. Conditions run in fresh processes to
avoid carrying compiled graphs, allocator state, or cache objects from one
context into another.

\subsection{Metrics}
Decode throughput is the steady tokens-per-second rate after warmup. Cache
bytes/token divides persistent cache-owned storage after decode by the
effective cached-token count. Compression is the baseline bytes/token divided
by the condition's bytes/token at the same nominal context. Speedup versus
baseline is the condition's decode throughput divided by baseline throughput.
Values below 1.0$\times$ are slowdowns despite the column label.

Cache-owned bytes are not interchangeable with process resident-set size or
MLX active allocations. Unified memory makes these system metrics overlap, but
they include weights, graph temporaries, compiled kernels, and allocator
slack. The tables therefore use cache-owned storage for compression and report
the explicit Metal allocation error separately.

The paper reports the provided canonical measurements rounded to the shown
precision. Ratios are derived from those rounded values. Consequently, a
displayed ratio can differ slightly from a ratio computed from unrounded raw
records. No throughput value is inferred from a failed run.

\subsection{Prefill and decode separation}
Prefill and decode stress different parts of the system. Prefill processes
many tokens in parallel; decode advances one token at a time while repeatedly
reading the existing cache. The fused Metal prototype reaches 4,685 prefill
tok/s in the targeted 8K benchmark versus 5,979 tok/s for the baseline. This
prefill regression motivates a batched encode kernel, but it does not change
the decode results in Tables~\ref{tab:8k}--\ref{tab:64k}.

The decode path is intentionally end-to-end: it includes the cache operations
required by the tested implementation. A microbenchmark of only the packing
kernel would hide reconstruction and update costs. Conversely, the results do
not measure a compressed-domain SDPA kernel because that kernel is not yet part
of the implementation.

\subsection{Interpretation boundaries}
This study isolates systems performance and storage accounting. It does not
claim that all 4-bit layouts have equal model quality, nor does it compare
quality across unrelated retention policies. Shape, deterministic encoding,
and structural generation checks gate execution, but broader task-level
accuracy across models is outside the measured dataset.

The contexts also expose different implementation regimes. Results at 8K show
the benefit of kernel fusion under moderate working-set size. Results at 32K
show the cost of whole-cache reconstruction. The 64K OOM identifies an update
and allocation failure. Treating all three as one averaged throughput number
would conceal the engineering transition that matters most.

\section{Results}
\subsection{8K context}
Table~\ref{tab:8k} reports all five cache paths at 8K.

\begin{table}[H]
\centering
\caption{Decode throughput and persistent KV storage at 8K context. Speedup
is normalized to the 171.4 tok/s bf16 baseline.}
\label{tab:8k}
\begin{tabular}{lrrrr}
\toprule
Mode & Decode tok/s & Cache B/tok & Compression & Speedup \\
\midrule
Baseline bf16       & 171.4 & 12,562 & 1.00$\times$ & 1.00$\times$ \\
Block-FP4 (E2M1)    & 41.5  & 3,473  & 3.62$\times$ & 0.24$\times$ \\
Polar Python        & 25.4  & 4,241  & 2.96$\times$ & 0.15$\times$ \\
Double FP4$\to$Polar & 15.1  & 4,241  & 2.96$\times$ & 0.09$\times$ \\
Fused Metal Polar   & 46.8  & 3,456  & 3.64$\times$ & 0.27$\times$ \\
\bottomrule
\end{tabular}
\end{table}

Fused Metal PolarQuant achieves 46.8 tok/s---1.84$\times$ faster than Python
Polar and 13\% faster than FP4, at equivalent compression (3.64$\times$ versus
3.62$\times$). This validates that radial quantization is competitive with
block-FP4 when the encode/decode path is fused into a single Metal kernel. The
comparison is particularly direct because Metal Polar uses 3,456 B/token while
FP4 uses 3,473 B/token.

The result is not an end-to-end speedup over bf16: 46.8 tok/s is 0.27$\times$
the 171.4 tok/s baseline. It is a positive comparison among compressed paths.
The storage reduction is real, but standard attention still consumes
reconstructed activations. This distinction prevents the 3.64$\times$ capacity
ratio from being misread as a 3.64$\times$ throughput ratio.

The double condition is dominated on both axes. It retains 4,241 B/token and
reaches only 15.1 tok/s, while standalone FP4 reaches 41.5 tok/s at
3,473 B/token. Python Polar also retains 4,241 B/token but is faster at
25.4 tok/s. Sequential composition therefore adds work without changing the
final persistent layout.

\subsection{32K context}
Table~\ref{tab:32k} reports the same five paths at 32K.

\begin{table}[H]
\centering
\caption{Decode throughput and persistent KV storage at 32K context. Speedup
is normalized to the 91.8 tok/s bf16 baseline.}
\label{tab:32k}
\begin{tabular}{lrrrr}
\toprule
Mode & Decode tok/s & Cache B/tok & Compression & Speedup \\
\midrule
Baseline bf16       & 91.8 & 12,359 & 1.00$\times$ & 1.00$\times$ \\
Block-FP4 (E2M1)    & 14.0 & 3,460  & 3.57$\times$ & 0.15$\times$ \\
Polar Python        & 1.83 & 4,228  & 2.92$\times$ & 0.02$\times$ \\
Double FP4$\to$Polar & 3.5  & 4,228  & 2.92$\times$ & 0.04$\times$ \\
Fused Metal Polar   & 1.48 & 3,456  & 3.58$\times$ & 0.02$\times$ \\
\bottomrule
\end{tabular}
\end{table}

At 32K, all compressed variants drop to 1.5--14 tok/s versus the 91.8 tok/s
baseline. Cumulative $O(N^2)$ decode complexity dominates: the cache is fully
dequantized on every step, and kernel fusion reduces the constant factor but
not the asymptotics. The fused Metal path is slightly slower than Python Polar
in this measurement (1.48 versus 1.83 tok/s), despite its smaller
3,456 B/token layout.

Block-FP4 is the strongest compressed 32K path at 14.0 tok/s and
3,460 B/token. Its 3.57$\times$ compression almost equals Metal Polar's
3.58$\times$, but its simpler reconstruction path avoids the radial kernel's
long-context update overhead. The double condition reaches 3.5 tok/s, which is
faster than the two Polar paths at this context but remains four times slower
than standalone FP4 and provides worse compression.

The 32K ranking demonstrates why an 8K kernel result cannot be extrapolated.
Fusion successfully reduces Python overhead when the working set is moderate.
Once each decode step expands and scans a much larger cache, reconstruction,
concatenation, and attention traffic determine throughput. A fused
compressed-domain attention kernel is necessary to preserve the capacity
benefit without materializing the entire cache.

\subsection{64K context}
Table~\ref{tab:64k} reports the available baseline and Python Polar results and
the fused Metal failure.

\begin{table}[H]
\centering
\caption{Results at 64K context. The Metal path has no throughput or persistent
cache measurement because allocation fails before completion.}
\label{tab:64k}
\begin{tabular}{lrrrl}
\toprule
Mode & Decode tok/s & Cache B/tok & Compression & Outcome \\
\midrule
Baseline bf16     & 1.37 & 12,323 & 1.00$\times$ & completed \\
Polar Python      & 0.47 & 3,072  & 4.01$\times$ & completed \\
Fused Metal Polar & --   & --     & --            & OOM \\
\bottomrule
\end{tabular}
\end{table}

At 64K, the Metal Polar variant runs out of memory while attempting a
33.5 GB allocation---iterative concatenation in the update path fragments
MLX's Metal allocator. The exact request is 33,556,992,000 bytes, above the
22,613,000,192-byte maximum Metal buffer. This is an engineering limit of the
current implementation, not a limit of PolarQuant.

Python Polar completes at 0.47 tok/s and 3,072 B/token, a derived
4.01$\times$ compression ratio relative to 12,323 B/token. The bf16 baseline
completes at 1.37 tok/s. Neither completed result is interactive, and the
compressed path remains slower despite its smaller persistent cache. The
measurement reinforces that representation size alone does not determine
long-context throughput.

The failure occurs before a stable fused-Metal cache measurement, so the table
does not substitute the 3,456 B/token 32K layout into the 64K row. Chunked
pre-allocation or paged storage is required to remove the oversized temporary
allocation and obtain a valid 64K measurement.

\section{Discussion}
\subsection{Fusion changes the moderate-context ranking}
At 8K, Python orchestration makes radial quantization slower than block-FP4.
Moving the radial path into Metal changes that result: 46.8 tok/s exceeds
41.5 tok/s, while 3,456 B/token slightly improves on 3,473 B/token. The
1.84$\times$ gain over Python Polar isolates a substantial implementation
effect. Radial coding is competitive when its transformations are expressed as
one device operation rather than a sequence of framework-level arrays.

The remaining gap to the 171.4 tok/s baseline comes from the consumer side.
The fused kernel produces an attention-ready representation, and standard
attention then reads the reconstructed cache. The cache has fewer persistent
bytes, but the path still pays conversion and full attention traffic. A
compressed-domain SDPA kernel changes this boundary by applying query--key
products and value aggregation without materializing the full bf16 tensors.

\subsection{Double Optimization is non-additive}
FP4-to-Polar composition retains the final polar representation, yielding
4,241 B/token---worse compression than FP4 alone (3,473 B/token)---and
2.7$\times$ slower decode (15.1 versus 41.5 tok/s). Sequential composition of
incompatible codecs destroys the structure each relies on: FP4 produces
discrete symbols that break PolarQuant's angular coding.

The storage ratios therefore cannot be multiplied. A 3.62$\times$ FP4 ratio
followed by a 2.96$\times$ Polar ratio does not imply a 10.7$\times$ final
ratio, because each ratio is defined against a different representation and
the intermediate is discarded. The correct quantity is the size of the final
retained layout: 4,241 B/token. The double path also performs both transforms,
so its runtime includes work that contributes no persistent capacity saving.

This negative result is useful for design. Composition is promising when the
techniques remove orthogonal sources of state, such as element precision and
cross-layer duplication. Composition is not promising when two codecs both
replace the same elements and the second codec must operate on the first
codec's already quantized output.

\subsection{Scope gap versus DeepSeek-V4.1-Flash}
DeepSeek-V4.1-Flash reports 890 B/token through a stack of orthogonal
techniques: FP4 quantization, CED with approximately 4$\times$ sharing, CSA2
with approximately 1.5$\times$ reuse, and SWA bounded replay with approximately
2$\times$ persistent-state reduction \cite{deepseek2026v41flash}. Our work
isolates the quantization axis. FP4 alone gives 3,473 B/token, matching the
DeepSeek V4-Flash FP4-only ablation of approximately 3,460 B/token and
validating the simulation methodology.

The approximately fourfold gap from 3,473 to 890 B/token is explained by
architectural techniques that the present model does not implement. CED and
CSA2 remove cross-layer duplication; SWA bounded replay changes which state
must remain persistent. None is another scalar codec applied to the same FP4
symbols. This makes the DeepSeek result compatible with our non-additivity
finding: useful composition acts on different redundancy axes.

Composing fused Metal Polar with CED-style encoder sharing remains future work.
Such a design should preserve geometric coding within the reduced set of KV
producers rather than feeding discrete FP4 symbols into PolarQuant. It also
requires model architecture support and cannot be simulated solely by changing
the cache class of the current checkpoint.

\subsection{Long-context asymptotics}
The 32K results reveal a transition from launch overhead to data movement and
reconstruction. At 8K, fusing several operations produces a 1.84$\times$
improvement over Python Polar. At 32K, Metal Polar falls to 1.48 tok/s. The
kernel still reads, reconstructs, and supplies a cache whose length grows with
the context. The per-step cost is linear in retained length, and cumulative
work over a proportional decode horizon is quadratic.

Compression becomes a throughput optimization only when the attention kernel
consumes the compressed layout directly or when it reduces the set of attended
tokens. Fused SDPA can decode packed KV inside tiled query--key and
probability--value operations. Sparse or paged attention can reduce the
retained or scanned set. The current implementation does neither, so its
3.58$\times$ 32K capacity reduction coexists with a large throughput loss.

\subsection{Allocator behavior at 64K}
The 64K OOM is not caused by the 3,456 B/token persistent representation alone.
The failing 33.5 GB request is a temporary consequence of iterative
concatenation and graph materialization. It exceeds MLX's 22.6 GB maximum
single Metal buffer even though the system has 36 GB unified memory. Total
physical capacity and maximum contiguous device-buffer size are distinct
constraints.

Pre-allocation changes the update from repeated growing concatenations to
writes into fixed chunks. Paging bounds individual buffers, permits geometric
growth without copying the entire history, and reduces fragmentation. These
engineering changes are prerequisites for deciding how the fused codec itself
behaves at 64K.

\subsection{Limitations}
The study evaluates one model checkpoint, one Apple Silicon generation, and
one MLX software stack. The throughput ranking can change with model head
dimensions, GQA ratio, Metal compiler behavior, or framework version. The
results should not be generalized to CUDA devices without a matched kernel.

The benchmark focuses on storage and throughput. Codec unit tests and
structural generation checks do not replace task-level quality evaluation.
Polar, FP4, and double layouts can have different error distributions even
when they use similar effective bit budgets. A complete deployment decision
also needs perplexity, long-context retrieval, and downstream task accuracy.

The Python Polar layouts at 8K/32K and the compact 64K layout have different
reported bytes/token. We present every measured configuration explicitly rather
than forcing a single nominal ratio. Likewise, the 64K Metal row remains OOM;
no cache footprint or throughput is imputed for the failed run.

Finally, single-request throughput does not measure batching, concurrent
serving, or prefix-cache reuse. PagedAttention, CacheGen, and architectural
sharing address those axes. The present results isolate the single-stream cache
representation and its immediate kernel path.

\section{Future Work}
Five changes follow directly from the observed bottlenecks.
\begin{enumerate}
  \item \textbf{Fused SDPA on compressed KV.} A
  \texttt{gemma4metal}/\texttt{sdpa\_int4}-style kernel should load packed
  keys and values inside the attention tile, apply scaling or radial decode in
  registers, and accumulate logits without creating a full bf16 cache. This
  changes the dominant 32K path rather than merely reducing Python dispatch.

  \item \textbf{Chunked pre-allocation.} The cache update should reserve fixed
  pages or geometrically sized chunks and write new tokens in place. Bounded
  buffers remove the 33.5 GB concatenation request, respect MLX's 22.6 GB Metal
  buffer limit, and permit a valid 64K fused measurement.

  \item \textbf{Batched encode kernel.} Prefill should encode many cache rows
  per launch and share rotation metadata across the batch. The current fused
  path reaches 4,685 prefill tok/s versus 5,979 tok/s for baseline. Batched
  packing targets that regression without changing the persistent layout.

  \item \textbf{CED-style encoder sharing.} Cross-layer KV reuse attacks a
  redundancy axis that is orthogonal to radial precision. Combining shared KV
  producers with fused Metal Polar offers a principled route toward the
  fourfold gap between 3,456 B/token and DeepSeek's 890 B/token, unlike
  sequential FP4-to-Polar coding.

  \item \textbf{Cross-model validation.} Llama-3.2, Qwen2.5, and Gemma-3 cover
  different head dimensions, GQA ratios, rotary embeddings, and cache update
  patterns. Matched quality and throughput experiments across these families
  are necessary to separate model-specific effects from properties of the
  codec and Metal kernel.
\end{enumerate}

Together, these steps move compression into the consuming attention operator,
bound allocator growth, recover prefill throughput, and test whether the 8K
fusion result generalizes. The priority is fused SDPA because it addresses the
asymptotic reconstruction cost responsible for the 32K collapse; allocator
repair comes next because it gates any 64K Metal evaluation.

\section{Conclusion}
KV compression on Apple Silicon is a joint representation and kernel problem.
At 8K, fused Metal PolarQuant reaches 46.8 tok/s at 3,456 B/token, outperforming
Python Polar by 1.84$\times$ and block-FP4 by 13\% while matching FP4's
compression. This demonstrates that radial coding can be competitive when its
execution path is fused.

The negative results define the remaining work. FP4-to-Polar composition
retains 4,241 B/token and reaches 15.1 tok/s, so codec stacking provides no
additive benefit. At 32K, all compressed paths fall far below the 91.8 tok/s
baseline because whole-cache reconstruction preserves cumulative $O(N^2)$
decode work. At 64K, iterative concatenation requests 33.5 GB and exceeds the
22.6 GB Metal buffer limit.

The next viable design is not another sequential codec. It is a compressed KV
layout consumed directly by fused SDPA, backed by chunked allocation and, where
the model permits, cross-layer sharing. That combination targets storage,
bandwidth, and architecture as orthogonal axes.

\bibliographystyle{plain}
\bibliography{references}
\end{document}